\documentstyle[twoside,fleqn,npb,epsfig]{article}
\newcommand{\AmS}{{\protect\the\textfont2
  A\kern-.1667em\lower.5ex\hbox{M}\kern-.125emS}}

\title{Parton Distributions
in the Virtual Photon Target\\
and Factorization Scheme Dependence\thanks{Talk given by
K. Sasaki at the Workshop 
              ``Loops and Legs in Quantum Field Theory'', Germany, April 2000}}

\author{Ken Sasaki\address{Department of Physics,
         Faculty of Engineering, Yokohama National University \\
        Yokohama 240-8501, Japan}%
        ~and
        Tsuneo Uematsu\address{Department of Fundamental Sciences,
        FIHS, Kyoto University, \\
        Kyoto 606-8501, Japan}}

\begin{document}

\begin{abstract}
We investigate parton distributions in the virtual photon target,
both polarized and unpolarized, up to the next-leading order (NLO) in QCD.
Parton distributions can be predicted completely up to NLO, but
they are factorization-scheme-dependent. We analyze parton distributions
in several factorization schemes and discuss their scheme dependence.
Particular attentions are paid to the axial anomaly effect on the first
moments of the polarized quark parton distributions, and also
to the large-$x$ behaviors of polarized and unpolarized parton
distributions.
\end{abstract}

\maketitle

\begin{picture}(5,2)(-330,-380)
\put(2.3,-75){YNU-HEPTh-00-101}
\put(2.3,-90){KUCP-156}
\put(2.3,-105){May 2000}
\end{picture}

\section{INTRODUCTION}

In $e^+ e^-$ collision experiments,
we can measure the spin-independent and spin-dependent
structure functions, $F_2^{\gamma}(x, Q^2, P^2)$ and
$g_1^{\gamma}(x, Q^2, P^2)$, of the  virtual photon
(Fig.1).
\vspace*{-0.3cm}
\begin{figure}[htb]
\hspace{1cm}
\includegraphics[width=5cm,height=5cm,keepaspectratio]{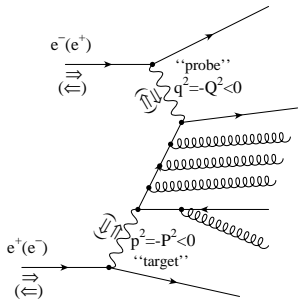}
\vspace{-1.8cm}
\caption{Deep inelastic scattering on a virtual photon in
$e^+ e^-$ collision.}
\label{fig:e^+ e^-}
\end{figure}

\vspace*{-0.8cm}

The advantage in  studying the virtual photon target is that, in the case
\begin{equation}
\Lambda^2 \ll P^2 \ll Q^2
\label{kin}
\end{equation}
where $-Q^2$ ($-P^2$) is the mass squared of
the probe (target) photon, and $\Lambda$ is the QCD scale parameter, we can
calculate the whole structure function up to the next-to-leading order (NLO)
by the perturbative method, in
contrast to the case  of the real photon target where in  NLO there exist
non-perturbative pieces. The NLO analyses of the virtual photon
structure functions,
$F_2^{\gamma}$ and $g_1^{\gamma}$,  have been  made by  Uematsu and Walsh
~\cite{UW} and the present authors~\cite{SU1}, respectively. In this talk
we will report the result of our investigation of the parton
distribution functions (pdf's) in the virtual photon target.
The  behaviors of the pdf's can be predicted entirely up to  NLO,
but they are factorization-scheme-dependent.
We carry out our analysis for the pdf's  of polarized and
unpolarized virtual photon
in several different factorization schemes, and see  how the pdf's change in
each
scheme.

\section{PARTON DISTRIBUTIONS IN VIRTUAL PHOTON}

We write down below the expressions for the polarized case.
The expressions for the unpolarized case are easily obtained
from the corresponding ones in the polarized case
by  removing the symble $\Delta$ and replacing $g_1^{\gamma}$
with $F_2^{\gamma}/x$.
Let $\Delta q^{\gamma}_S$($\Delta q^{\gamma}_{NS}$),
$\Delta G^{\gamma}$,
$\Delta \Gamma^{\gamma}$ be the flavor singlet
(non-singlet)-quark,  gluon, and  photon distribution functions, respectively,
in the longitudinally polarized virtual photon with mass $-P^2$.
In the leading order of the electromagnetic coupling constant,
$\alpha=e^2/4\pi$,
$\Delta \Gamma^{\gamma}$ does not evolve with $Q^2$ and is set to be
$\Delta \Gamma^{\gamma}=\delta(1-x)$.
In terms of the Mellin moments of these pdf's, the moment of the polarized
virtual photon  structure function
$g_1^{\gamma}(x, Q^2, P^2)$ is expressed in the QCD improved parton model as
\begin{equation}
 g_1^{\gamma}(n,Q^2,P^2)=\Delta{\mbox{\boldmath $C$}}^{\gamma}(n,Q^2) \cdot
\Delta{\mbox{\boldmath $q$}}^{\gamma}(n,Q^2,P^2)
\label{gonegamma} \\
\end{equation}
where
\begin{eqnarray}
&&\hspace{-0.6cm}
\Delta{\mbox{\boldmath $C$}}^{\gamma}(n,Q^2)=(\Delta C^{\gamma}_S~,~
 \Delta C^{\gamma}_G~,~\Delta C^{\gamma}_{NS}~,~
  \Delta C^{\gamma}_{\gamma}) \nonumber  \\
&&\hspace{-0.6cm}
\Delta{\mbox{\boldmath $q$}}^{\gamma}(n,Q^2,P^2)=(\Delta q_S^{\gamma}~,~
\Delta G^{\gamma}~,~ \Delta q_{NS}^{\gamma}~, ~\Delta \Gamma^{\gamma})
\nonumber
\end{eqnarray}
and $\Delta C^{\gamma}_S$($\Delta C^{\gamma}_{NS}$), $\Delta
C^{\gamma}_G$, and
$\Delta C^{\gamma}_{\gamma}$ are the moments of the coefficient functions
corresponding to singlet(non-singlet)-quark, gluon, and photon, respectively,
and they are independent of $P^2$.

The pdf's $\Delta{\mbox{\boldmath $q$}}^{\gamma}$ satisfy  inhomogeneous
evolution equations. The explicit expressions of $\Delta q^\gamma_S$,
$\Delta G^\gamma$, and $\Delta q^\gamma_{NS}$ up to the NLO are derived
from Eq.(4.46) of Ref.\cite{SU1}.
They are given in
terms of  one-(two-) loop hadronic anomalous dimensions
$\Delta\gamma_{ij}^{(0),n}$
($\Delta\gamma_{ij}^{(1),n}$) ($i,j=\psi, G$) and
$\Delta\gamma_{NS}^{(0),n}$
($\Delta\gamma_{NS}^{(1),n}$), one-(two-) loop anomalous dimensions
$\Delta K_i^{(0),n}$ ($\Delta K_i^{(1),n}$) ($i=\psi, G, NS$) which
represent the
mixing between photon and three hadronic operators $R_i^n$ ($i=\psi, G,
NS$), and
finally
$\Delta A_i^n$, the one-loop photon matrix elements  of  these
hadronic operators renormalized at $\mu^2=P^2(=-p^2)$,
\begin{equation}
  \langle \gamma (p) \mid R_i^n (\mu) \mid \gamma (p) \rangle \vert_{\mu^2=
P^2}
=\frac{\alpha}{4\pi} \Delta  A_i^n~.
\label{Initial}
\end{equation}

\section{FACTORIZATION SCHEMES}

Although $g_1^{\gamma}$ is a
physical quantity and thus unique, there remains a freedom in the
factorization of
$g_1^{\gamma}$  into $\Delta{\mbox{\boldmath $C$}}^{\gamma}$ and
$\Delta{\mbox{\boldmath $q$}}^{\gamma}$. Given the formula
Eq.(\ref{gonegamma}),
we can always redefine $\Delta{\mbox{\boldmath $C$}}^{\gamma}$ and
$\Delta{\mbox{\boldmath $q$}}^{\gamma}$ as follows
\cite{FP}:
\begin{eqnarray}
\Delta{\mbox{\boldmath $C$}}^{\gamma}(n,Q^2)
& \hspace{-0.3cm}\rightarrow\hspace{-0.3cm}&
\Delta{{\mbox{\boldmath $C$}}}^{\gamma}(n,Q^2)\vert_a \nonumber \\
     && \equiv\Delta{\mbox{\boldmath $C$}}^{\gamma}(n,Q^2)Z^{-1}_a(n,Q^2)
 \nonumber \\
\Delta{\mbox{\boldmath $q$}}^{\gamma}(n,Q^2,P^2)
&\hspace{-0.3cm}\rightarrow\hspace{-0.3cm}&
\Delta{{\mbox{\boldmath $q$}}}(n,Q^2,P^2)\vert_a \nonumber \\
 &&\equiv Z_a(n,Q^2)~ \Delta{\mbox{\boldmath $q$}}^{\gamma}(n,Q^2,P^2)
 \nonumber
\end{eqnarray}
where $\Delta{{\mbox{\boldmath $C$}}}^{\gamma}\vert_a  $ and $ \Delta
{{\mbox{\boldmath $q$}}}\vert_a $  correspond to the quantities in a new
factorization scheme-$a$. The most general form of a transformation for the
coefficient functions  in one-loop order, from $\overline {\rm MS}$ scheme to
a new factorization scheme-$a$, is given by
\begin{eqnarray}
    \Delta C_{S,~a}^{\gamma,~ n}&=&
\Delta C_{S,~\overline{\rm MS}}^{\gamma,~ n}~
-\langle e^2\rangle \frac{\alpha_s}{2\pi}~\Delta w_a(n) \nonumber \\
\Delta C_{G,~a}^{\gamma,~ n}&=&\Delta C_{G,~\overline {\rm MS}}^{\gamma,~ n}~
-\langle e^2\rangle \frac{\alpha_s}{2\pi}~\Delta z_a(n)  \nonumber\\
\Delta C_{NS,~a}^{\gamma,~ n}&=&\Delta C_{NS,~\overline {\rm MS}}^{\gamma,~ n}~
-\frac{\alpha_s}{2\pi}~\Delta w_a(n) \label{Coeffgamma} \\
 \Delta C_{\gamma,~a}^{\gamma,~ n}&=&
\Delta C_{\gamma,~\overline{\rm MS}}^{\gamma,~ n}~
-\frac{\alpha}{\pi}~3\langle e^4\rangle \Delta {\hat z}_a(n) \nonumber
\end{eqnarray}
where  $\langle e^2\rangle =\sum_i e^2_i/N_f$, $\langle e^4\rangle
=\sum_i e^4_i/N_f$, with
$N_f$ being the number of flavors of active quarks and $e_i$ being the
electric charge of $i$-flavor-quark.

Once the relations (\ref{Coeffgamma}) between the coefficient functions in the
$a$-scheme and $\overline {\rm MS}$ scheme are given, we can derive
corresponding
transformation rules~\cite{SU2} from $\overline {\rm MS}$ scheme to $a$-scheme
for the relevant two-loop anomalous dimensions and also
for the one-loop photon matrix elements,
$\Delta A_{\psi}^n$ and $\Delta A_{NS}^n$, of the quark operators.
Note that, in one-loop order, the photon matrix elements of gluonic
operators $R_G^n$ vanish in any scheme,  $\Delta A_{G}^n =0$.

We consider three different factorization schemes both in the polarized and
unpolarized cases.

\subsection{The polarized case}
\vspace{0.1cm}

(i) [The $\overline {\rm MS}$ scheme]
\ \ This is the only scheme in which both relevant one-loop coefficient
functions and
two-loop anomalous dimensions~\cite{MvN,V} were actually
calculated.  In the  $\overline {\rm
MS}$ scheme, the QCD (QED)  axial anomaly
resides  in the quark distributions and not in the gluon (photon)
coefficient function\cite{BQ,HYC}.
In fact we observe
\begin{eqnarray}
\Delta \gamma^{(1),n=1}_{\psi\psi,~\overline {\rm MS}}&=&
24C_FT_f \neq 0~, \nonumber \\
\Delta B_{G,~\overline {\rm MS}}^{n=1}&=&\Delta B_{\gamma,~
\overline{\rm MS}}^{n=1}=0~.
\end{eqnarray}
Also the first moment of the one-loop photon matrix element of quark operators
gains the non-zero values, i.e.,
\begin{eqnarray}
\Delta A_{\psi,~\overline {\rm MS}}^{n=1}&=&\frac{\langle e^2\rangle
}{\langle e^4\rangle -\langle e^2\rangle ^2}~ \Delta A_{NS,~\overline {\rm
MS}}^{n=1}
\nonumber\\ &=&-12\langle e^2\rangle N_f  \label{PhotonMSn=1}
\end{eqnarray}
which is due to the QED axial anomaly.

(ii) [The chirally invariant (CI)  scheme]
\ \ In this scheme the factorization of the photon-gluon (photon-photon) cross
section into the hard and soft parts is made so that  chiral symmetry is
respected and
all the anomaly effects are absorbed into the gluon (photon) coefficient
function\cite{HYC,MT}. Thus the spin-dependent quark distributions in
the CI
scheme are  anomaly-free. In particular, we have
\begin{eqnarray}
 \Delta B_{G,~{\rm CI}}^{n=1}&=&-2N_f~, \quad  \Delta B_{\gamma,~{\rm
CI}}^{n=1}=-4 \label{CIlike}\\
\Delta \gamma^{(1),n=1}_{\psi\psi,~{\rm CI}}&=&0 ~, \quad
\Delta A_{\psi,~{\rm CI}}^{n=1}= \Delta A_{NS,~{\rm CI}}^{n=1}=0~. \nonumber
\end{eqnarray}
The transformation from the
$\overline {\rm MS}$ scheme to the CI scheme is achieved by
\begin{eqnarray}
  \Delta w_{{\rm CI}}(n)&=&0~,\nonumber  \\
  \Delta  z_{{\rm CI}}(n)&=&\Delta  {\hat z}_{{\rm CI}}(n)
=2N_f\frac{1}{n(n+1)}~.  \label{TransCI}
\end{eqnarray}

(iii) [The off-shell (OS)  scheme]
\ \ In this scheme~\cite{BFR} we renormalize operators while keeping the
incoming
particle
off-shell, $p^2\neq 0$, so that at renormalization (factorization) point
$\mu^2=-p^2$, the finite terms vanish.  This is exactly the same as ``the
momentum
subtraction scheme" which was used some time ago to calculate, for instance,
the polarized quark and gluon coefficient  functions~\cite{KMSU,JK}.
The CI-relations in Eq.(\ref{CIlike}) also hold in the OS scheme.  The
transformation  from
$\overline{\rm MS}$  to the OS scheme is made by choosing
\begin{eqnarray}
&&\hspace{-0.6cm}\Delta w_{{\rm OS}}(n) \nonumber\\
&&\hspace{-0.6cm}= C_F \biggl\{\Bigl[S_1(n)\Bigr]^2+3S_2(n)-
S_1(n)\Bigl(\frac{1}{n}- \frac{1}{(n+1)}  \Bigr)  \nonumber  \\
&&\hspace{-0.6cm}\quad -\frac{7}{2} +\frac{2}{n}
-\frac{3}{n+1}-\frac{1}{n^2}+\frac{2}{(n+1)^2}\biggr\} \\
&&\hspace{-0.6cm}
\Delta  z_{{\rm OS}}(n)=\Delta  {\hat z}_{{\rm OS}}(n) \nonumber\\
&&\hspace{-0.6cm}
=N_f\biggl\{ -\frac{n-1}{n(n+1)}S_1(n)+ \frac{1}{n}
+\frac{1}{n^2}-\frac{4}{(n+1)^2} \biggr\}~.\nonumber
\end{eqnarray}
It is noted that in the OS scheme we have
$\Delta A^n_{\psi,~{\rm OS}}=\Delta A^n_{NS,~{\rm OS}}=0 $
for all $n$.

\subsection{The unpolarized case}
\vspace{0.1cm}

To study the pdf's inside unpolarized virtual photon,
we consider three factorization schemes:
(i) The $\overline {\rm MS}$ scheme;
(ii) The off-shell (OS)  scheme; and
(iii) The ${\rm DIS}_{\gamma}$ scheme.
The transformation from
$\overline {\rm MS}$ to the OS scheme is achieved by
\begin{eqnarray}
   &&\hspace{-0.6cm} w_{{\rm OS}}(n) =\Delta w_{{\rm OS}}(n) \nonumber \\
&&\hspace{-0.6cm}
z_{{\rm OS}}(n) ={\hat z}_{{\rm OS}}(n)=N_f\biggl\{
-\frac{n^2+n+2}{n(n+1)(n+2)}S_1(n)\nonumber\\
&& \qquad  + \frac{1}{n}
-\frac{1}{n^2} +\frac{4}{(n+1)^2}-\frac{4}{(n+2)^2} \biggr\}~.
\end{eqnarray}

The ${\rm DIS}_{\gamma}$
was introduced~\cite{GRV} some time ago for the analysis of the unpolarized
real photon structure function $F_2^{\gamma}(x, Q^2)$ in NLO. In this scheme
the direct-photon contribution to $F_2^{\gamma}$ is absorbed into the photonic
quark distributions, so that we take
\begin{eqnarray}
&&w_{{\rm DIS}_{\gamma}}(n)=
z_{{\rm DIS}_{\gamma}}(n)=0  \nonumber \\
&&{\hat z}_{{\rm DIS}_{\gamma}}(n)= \frac{N_f}{4}
B_{\gamma,~\overline {\rm MS}}^n  \\
&&\quad =N_f\biggl\{ -\frac{n^2+n+2}{n(n+1)(n+2)}S_1(n)\nonumber \\
&&\qquad \qquad -\frac{1}{n}+\frac{1}{n^2} +\frac{6}{n+1}- \frac{6}{n+2}
\biggr\}\nonumber
\end{eqnarray}

With these preparations, we now examine the factorization scheme
dependence of the pdf's in virtual photon.

\section{THE $n=1$ MOMENTS OF POLARIZED PDF'S}

The first moments of polarized pdf's are particularly interesting
due to their relevance to the axial anomaly~\cite{JK}.
>From now on we omit to write the explicit $Q^2$- and $P^2$-dependeces in
$\Delta q_S^\gamma$, $\Delta q_{NS}^\gamma$, $\Delta G^{\gamma}$ and
in their unpolarized counterparts.  For the CI and OS factorization
schemes, we have
\begin{eqnarray}
   \Delta w_a(n=1)&=&0,~\nonumber \\
 \Delta z_a(n=1)&=&\Delta{\hat z}_a(n=1)=N_f
\end{eqnarray}
where $ a={\rm CI,OS}$.
These schemes, therefore, give the same first moments for the pdf's.
In fact, we find~
$ \Delta A_{\psi,~a}^{n=1}= \Delta A_{NS,~a}^{n=1}=0$~and
this leads to
\begin{eqnarray}
\Delta q_S^\gamma(n=1)\vert_a =\Delta
q_{NS}^\gamma(n=1)\vert_a
=0   \label{QSCI}
\end{eqnarray}
up to  NLO. In these schemes, the axial anomaly effects
are transfered to the gluon and photon coefficient functions.
On the other hand, in the
$\overline {\rm MS}$ scheme the axial anomaly effects are
retained in the quark distributions.
In fact we  obtain for singlet quark pdf, for example,
\begin{eqnarray}
&&  \Delta q_S^{\gamma}(n=1)\vert_{\overline {\rm MS}}
\nonumber \\
&&\quad =
\Bigl[-\frac{\alpha}{\pi}~3 \langle e^2\rangle N_f
\Bigr]\nonumber \\
&&\qquad \times
\left\{1-\frac{2}{\beta_0}\frac{\alpha_s(P^2)-\alpha_s(Q^2)}{\pi}
 N_f\right\}~.   \label{QSMS}
\end{eqnarray}
The factor $\Bigl[-\frac{\alpha}{\pi}~3
\langle e^2 \rangle N_f\Bigr] $ is
related to the QED axial anomaly and the term
$\frac{2}{\beta_0}\frac{\alpha_s(P^2)-\alpha_s(Q^2)}{\pi}
 N_f$ is coming from the QCD axial anomaly~\cite{SU3}.

For gluon distribution, we obtain in NLO
\begin{eqnarray}
&&\Delta G^\gamma(n=1)\nonumber  \\
&&\quad=\frac{12\alpha}{\pi\beta_0}
\langle e^2\rangle
N_f~\frac{\alpha_s(Q^2)-\alpha_s(P^2)}{\alpha_s(Q^2)}~,
\end{eqnarray}
the same result for $\overline {\rm MS}$, CI and OS schemes.

\section{The PDF'S NEAR $x=1$}

The behaviors of pdf's near~$x=1$
are governed by the large-$n$ limit of those moments.

\subsection{The polarized case}
\vspace{0.1cm}

The pdf's in LO are factorization-scheme independent. For large
$n$,~
$\Delta q_S^{\gamma}(n)\vert_{\rm LO}$ and $\Delta
q_{NS}^{\gamma}(n)\vert_{\rm LO}$ behave as $1/(n~{\rm ln}~n)$, while
$\Delta G^{\gamma}(n)\vert_{\rm LO}\propto 1/(n~{\rm ln}~n)^2$.
Thus in $x$ space,  the pdf's vanish for $x \rightarrow 1$.
In fact we find
\begin{eqnarray}
&& \Delta q_S^{\gamma}(x)\vert_{\rm LO}\nonumber  \\
&&\quad \approx
\frac{\alpha}{4\pi}\frac{4\pi}{\alpha_s(Q^2)}
N_f\langle e^2\rangle \frac{9}{4}~\frac{-1}{{\rm ln}~(1-x)} \\
&&\Delta G^{\gamma}(x)\vert_{\rm LO}\nonumber  \\
&&\quad \approx
\frac{\alpha}{4\pi}\frac{4\pi}{\alpha_s(Q^2)}
N_f\langle e^2\rangle \frac{1}{2}~\frac{-{\rm ln}~x}{{\rm
ln}^2~(1-x)}~.
\nonumber
\end{eqnarray}
The behaviors of $\Delta q_{NS}^{\gamma}(x)$ for $x \rightarrow 1$,
both in LO and NLO, are always
given by the corresponding expressions for $\Delta q_{S}^{\gamma}(x)$
with  replacement of the charge factor $\langle e^2\rangle $
with
$(\langle e^4\rangle -\langle e^2\rangle ^2)$.

{}From analysis of the large $n$ behaviors for
the moments of the NLO pdf's in the $\overline {\rm MS}$
scheme, we find near $x=1$,
\begin{eqnarray}
\Delta q_S^{\gamma}(x)\vert_{{\rm NLO},
\overline {\rm MS}}
&\approx&
\frac{\alpha}{4\pi} N_f\langle e^2\rangle 6~[-{\rm
ln}(1-x)], \nonumber \label{NLOMSbar}\\
\Delta G^{\gamma}(x)\vert_{{\rm NLO},\overline {\rm
MS}}
&\approx&
\frac{\alpha}{4\pi} N_f\langle e^2\rangle 3~[-{\rm
ln}~x] ~.
\end{eqnarray}
It is remarkable that in the $\overline {\rm MS}$ scheme quark
pdf's,
$\Delta q_S^{\gamma}(x)\vert_{{\rm NLO},\overline {\rm MS}}$ and
$\Delta q_{NS}^{\gamma}(x)\vert_{{\rm NLO},\overline {\rm MS}}$,
diverge as $[-{\rm
ln}(1-x)]$ for $x\rightarrow 1$.
The NLO quark pdf's in the CI scheme also diverge
as $x\rightarrow 1$. In fact we observe that
$\Delta q_S^{\gamma}(x)\vert_{{\rm NLO},{\rm CI}}$ approaches
$\Delta q_S^{\gamma}(x)\vert_{{\rm NLO},\overline {\rm MS}}$
for large $x$.


On the other hand, the OS scheme gives quite different behaviors near
$x=1$ for  the quark pdf's.
We find that, in $x$ space,
$\Delta q_{S}^\gamma(x)\vert_{{\rm NLO},{\rm OS}}$
does not diverge for $x \rightarrow 1$ but approaches a constant value:
\begin{eqnarray}
\Delta q_{S}^\gamma(x)\vert_{{\rm NLO},{\rm OS}}&\rightarrow&
\frac{\alpha}{4\pi}N_f \langle e^2\rangle
\Bigl[\frac{69}{8} + \frac{3}{4}N_f \Bigr]. \label{NLOOS}
\end{eqnarray}

\subsection{The unpolarized case}
\vspace{0.1cm}

In LO the pdf's of  unpolarized virtual photon target have the same behaviors
as the polarized case for  $x \rightarrow 1$. We obtain
\begin{eqnarray}
q_S^{\gamma}(x)\vert_{\rm LO} &\approx&\Delta q_S^{\gamma}
(x)\vert_{\rm LO} \\
G^{\gamma}(x)\vert_{\rm LO}  &\approx&
 \Delta G^{\gamma}(x)\vert_{\rm LO}
 \nonumber
\end{eqnarray}

Furthermore, we have found that the NLO behaviors of the pdf's, which are
predicted by each factorization scheme for $x\rightarrow 1$,
are the same both in the unpolarized and polarized cases. More
specifically
\begin{eqnarray}
q_S^{\gamma}(x)\vert_{{\rm NLO},~a} &\approx&
 \Delta q_S^{\gamma}(x)\vert_{{\rm NLO},~a} \\
G^{\gamma}(x)\vert_{{\rm NLO},~a}&\approx&
\Delta G^{\gamma}(x)\vert_{{\rm NLO},~a} \nonumber
\end{eqnarray}
where $a=\overline {\rm MS},~~{\rm OS}$.

In ${\rm DIS}_{\gamma}$ scheme, quark pdf's becomes negative and divergent for
$x \rightarrow 1$. In fact, we find
\begin{eqnarray}
 q_{S}^\gamma(x)\vert_{{\rm NLO},{\rm DIS}_{\gamma}}
\approx\frac{\alpha}{4\pi}~ 6N_f \langle e^2\rangle  [{\rm ln}~(1-x)]
\end{eqnarray}
This is due to the fact that the photonic coefficient function
$C_{\gamma}^{\gamma}(x)$, which becomes negative and divergent for
$x\rightarrow 1$ in $\overline {\rm MS}$, is absorbed into the quark pdf's in
the ${\rm DIS}_{\gamma}$ scheme.

\begin{figure}[htb]
\vspace{-1.8cm}
\includegraphics[width=8cm,height=10cm,keepaspectratio]{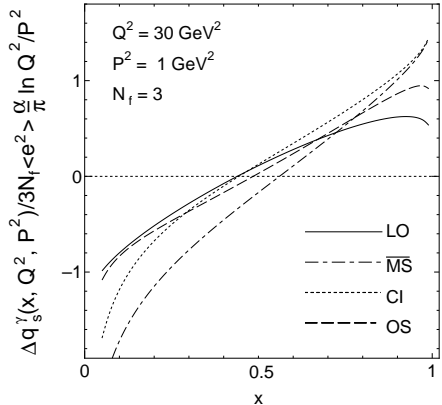}
\vspace{-1.8cm}
\caption{The polarized
singlet quark distribution
$\Delta q_S^{\gamma}(x,Q^2,P^2)$ in LO (solid line) and beyond LO.
The NLO results are from
$\overline {\rm MS}$(dash-dotted line), CI (short-dashed line),
and OS (long-dashed line) schemes.}
\label{fig2}
\end{figure}

\begin{figure}[htb]
\vspace{-1.8cm}
\includegraphics[width=8cm,height=10cm,keepaspectratio]{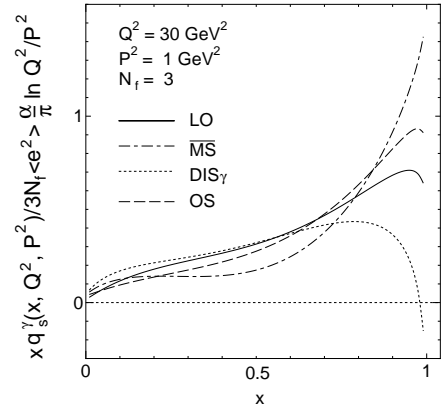}
\vspace{-1.8cm}
\caption{The unpolarized
singlet quark distribution
$xq_S^{\gamma}(x,Q^2,P^2)$ in LO (solid line) and beyond LO.
The NLO results are from  $\overline {\rm MS}$(dash-dotted line),
OS (long-dashed line), and  ${\rm DIS}_{\gamma}$(short-dashed line) schemes. }
\label{fig3}
\end{figure}

\section{NUMERICAL ANALYSIS}

The pdf's are recovered from the moments by the
inverse Mellin transformation. In Fig.2 we plot
the singlet quark pdf  $\Delta q_S^{\gamma}(x,Q^2,P^2)$
of polarized virtual photon both in LO and NLO
in units of $(3 N_f \langle e^2\rangle\alpha/\pi)~{\rm ln}(Q^2/P^2)$.
We have taken $N_f=3$, $Q^2=30~ {\rm GeV}^2$, $P^2=1~ {\rm GeV}^2$,  and the
QCD
scale parameter $\Lambda=0.2~ {\rm GeV}$.
We present the NLO results in three different factorization schemes,
i.e., $\overline {\rm MS}$, CI and OS.
The CI and OS lines cross the $x$-axis nearly at the same point, just below
$x=0.5$, while the $\overline {\rm MS}$ line crosses at above $x=0.5$.
This is understandable since we saw from Eq.(\ref{QSCI}) that the first moment
of $\Delta q_S^\gamma$ vanishes in the CI and OS schemes, while it is negative
in the $\overline {\rm MS}$ scheme. As $x\rightarrow 1$, we observe that the
$\overline{\rm MS}$ and  CI lines  continue to increase
and actually tend to merge,  while the OS line starts to drop. These behaviors
are inferred from Eqs.(\ref{NLOMSbar}-\ref{NLOOS}).
It is noted that the $\overline {\rm MS}$ line is much different from
the LO one. We see that OS scheme predicts a better behavour
for $\Delta q_S^{\gamma}$ than other schemes in the sence that
the OS line is closer to the LO one and does not diverge as $x\rightarrow 1$.

Concerning the non-singlet quark distribution
$\Delta q_{NS}^{\gamma}(x,Q^2,P^2)$, we find that when we take  into account
the charge factors, it falls on the singlet
quark distribution  in almost all $x$ region; namely two ``normalized"
distributions $\Delta {\widetilde q}_S^{\gamma}\equiv\Delta
q_S^{\gamma}/\langle
e^2\rangle $ and $\Delta{\widetilde q}_{NS}^{\gamma}\equiv \Delta
q_{NS}^{\gamma}/(\langle e^4\rangle -\langle e^2\rangle ^2)$  mostly overlap
except at very small $x$ region.  Finally, compared with quark pdf's, the
gluon
distribution $\Delta G^{\gamma}(x,Q^2,P^2)$
is very much small in absolute value  except at the small $x$ region.

In Fig.3 we plot
the singlet quark pdf  $x q_S^{\gamma}(x,Q^2,P^2)$
inside unpolarized virtual photon target both in LO and NLO
in units of $(3 N_f \langle e^2\rangle\alpha/\pi)~{\rm ln}(Q^2/P^2)$.
Again we have taken $N_f=3$, $Q^2=30~ {\rm GeV}^2$, $P^2=1~ {\rm GeV}^2$,  and
the QCD scale parameter $\Lambda=0.2~ {\rm GeV}$.
We present the NLO results in three different factorization schemes,
i.e., $\overline {\rm MS}$, OS and ${\rm DIS}_{\gamma}$.
As in the polarized case, the  $\overline {\rm MS}$ line deviates
from the LO one, and diverges as $x\rightarrow 1$.
The ${\rm DIS}_{\gamma}$ line is close to the LO line below $x<0.7$, but
negatively diverges as $x\rightarrow 1$. Again, in the unpolarized case,
the OS scheme gives a better behavior for $x q_S^{\gamma}$.
Actually, the OS line is closer to the LO one and starts to drop to reach
the finite value for  $x\rightarrow 1$.

\section{SUMMARY}

The behaviors of the pdf's inside the virtual photon target,
polarized and unpolarized, can be predicted entirely up to NLO,
but they are factorization-scheme-dependent. We have studied
the scheme dependence of the pdf's in the virtual photon. In the case of
polarized pdf's, the scheme dependence is clearly seen in the first
moments and the large-$x$ behaviors of quark distributions. In the
umpolarized case,  the scheme dependence is also observed in the
large-$x$ behaviors of quark distributions.
The NLO quark pdf's predicted by the $\overline {\rm MS}$
scheme deviate substantially from the LO results and
diverge as $x\rightarrow 1$, for both polarized and unpolarized cases.
On the other hand, the OS scheme gives better behaviors for the
quark pdf's in the sence that they are close to the LO pdf's
and remain finite as $x\rightarrow 1$.

\end{document}